\documentclass[prc,aps,superscriptaddress,showpacs]{revtex4}
\usepackage{epsfig}
\usepackage{graphicx}
\usepackage{amsmath}
\usepackage{yhmath}
\usepackage{amssymb}

\date{\today}

\tighten

\newcommand{\disregard}[1]{}

\newcommand{\be}{\begin{equation}}
\newcommand{\ee}{\end{equation}}
\newcommand{\bn}{\begin{eqnarray}}
\newcommand{\en}{\end{eqnarray}}
\newcommand{\ba}{\begin{array}}
\newcommand{\ea}{\end{array}}
\newcommand{\bc}{\begin{center}}
\newcommand{\ec}{\end{center}}
\newcommand{\bml}{\begin{mathletters}}
\newcommand{\eml}{\end{mathletters}}

\tighten

\begin{document}

\draft

\preprint{\fbox{\sc version of \today}}

\title{Pairing in nuclei}

\author{{\sc  Wojciech Satu{\l}a}}
\email{             satula@fuw.edu.pl}
\affiliation{Institute of Theoretical Physics, University of Warsaw,\\
             ul. Ho{\.z}a 69, PL-00 681 Warsaw, Poland}

\begin{abstract}
Simple generic aspects of nuclear pairing in homogeneous medium as well
as in finite nuclei are discussed. It is argued that low-energy
nuclear structure is not sensitive enough to resolve fine details
of nuclear nucleon-nucleon (NN) interaction in general and pairing
NN interaction in particular what allows for regularization of
the ultraviolet (high-momentum) divergences and a consistent formulation
of effective superfluid local theory. Some aspects of (dis)entanglement
of pairing with various other effects as well as forefront ideas
concerning isoscalar pairing  are also briefly discussed.
\end{abstract}


\maketitle

\section{Introduction}

The pioneering works of Bohr, Mottelson and Pines~\cite{[Boh58]},
Belyayev~\cite{[Bel59]} and, in particular, the ultimate success of
early "large-scale" BCS calculations by Nilsson and Prior~\cite{[Nil61]}
%invoking Nilsson hamiltonian augmented by monopole residual pairing
%interaction treated within the BCS approximation to
explaining simultaneously odd-even mass staggering and moments of
inertia neatly settled pairing just at the heart of nuclear physics.
The success of Nilsson-plus-BCS approach using a simple constant
matrix element pairing interaction confirms the absolute dominance
of the $s$-wave (monopole) part of the
in-medium effective particle-particle (p-p)
NN interaction at the Fermi energy.
There is, however, yet another very general argument pointing toward the
simplicity of nuclear pairing. Indeed, since pairing modifies the nucleonic
motion essentially only in the closest vicinity of the Fermi energy,
$E_F - \Delta \leq
\frac{(p_F\pm \delta p)^2}{2m} \leq E_F + \Delta$
it gives rise to uncertainty in momentum space,
$\delta p \sim \Delta/v_F$, what translates to uncertainty in coordinate
space of the order of
$\xi \sim \frac{(\hbar c)^2 k_F}{(mc^2)\Delta} \sim 50$\,fm.
The quantity $\xi$, which is known as the coherence length,
defines the spatial extension of the nucleonic Cooper pair. Note, that
[in fact, due to $E_F \gg \Delta$] the value of $\xi$ exceeds by far
the typical interaction range $\xi \gg r_o \sim \frac{1}{k_F}$ what is
known as weak coupling limit.

The above argumentation holds also for finite nuclei where
$\xi \sim 2R \gg r_o$ [$R$ denotes nuclear radius] i.e.
nucleonic Cooper pairs are spatially very extended
objects. They are therefore not bosons, they are overlapping
pairs which, for example, cannot form Bose-Einstein condensate.
Since nuclear pairing is characterized by a small parameter
$\epsilon \equiv \frac{r_o}{\xi} \ll 1$
it should posses an intrinsic simplicity. In
particular, it should be fairly insensitive to fine details of
the NN interaction or, in alternative words, should be well described
by a local theory.
%Clear signals of such an independence
%come from the $^1$S$_0$ pairing gap calculations in homogenous media.
In fact, since $r_o \ll 2R$, arguments speaking in favor of a local
approximation can be extended over to the particle-hole (p-h)
channel as well, see~\cite{[Sat05]} and refs. quoted therein.

The paper is organized as follows. In Sect.~\ref{1s0} I shall
discuss the independence of $^1$S$_0$ pairing  with respect to
details of the NN interaction
in homogeneous medium. In Sect.~\ref{local} I shall present possible
regularization schemes leading to a cutoff parameter independent
superfluid local density approximation. In Sect.~\ref{ent}
I shall briefly discuss the perplexing problem of (dis)entanglement between
pairing and various other effects. Finally, in Sect.~\ref{pnpair}
I shall briefly overview current ideas regarding isoscalar pairing.

\section{The $^1$S$_0$ pairing gap in infinite homogeneous medium}\label{1s0}

\begin{figure}[t]
\begin{center}
%\includegraphics[scale=0.70, angle=0.0, clip, viewport= 150 300 600
%670]{so-fig01.eps}
\includegraphics[scale=0.70, angle=0.0,clip]{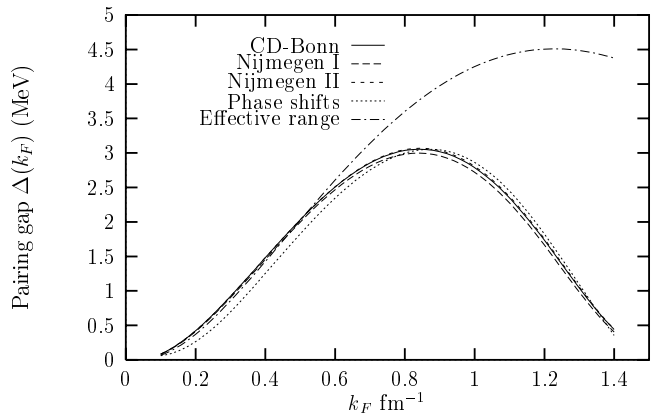}
\caption[]{$^1$S$_0$ pair-gap caculated using CD-Bonn and Nijmegen
bare NN interactions and free spectrum. Dotted line shows calculations
using phase-shift approximation. Dash-dotted line indicates results
obtained by using finite range approximation for phase shifts.
Taken from Ref.~\protect{\cite{[Elg98]}}.
}
\label{freegap}
\end{center}
\end{figure}

Several bare NN potentials are available nowadays that fit
the two-body scattering data with very high precision.
In spite of substantial differences among them they all predict
essentially the same values for the $^1$S$_0$ pair-gap
%$\Delta (k_F)$
%\be\label{delta}
%\Delta (k) = -\frac{1}{\pi} \int_0^\infty dk'k'^2 v(k,k')
%\frac{\Delta(k')}{E(k')}
%\ee [$E(k)$ denotes quasi-particle (QP) energy]
over rather wide range of
Fermi momenta $k_F$, see examples in Fig.~\ref{freegap}.
It appears that the detailed NN interaction is not needed
at all to determine the $^1$S$_0$ gap, as nicely demonstrated
in Ref.~\cite{[Elg98]}. The decisive point is
that the $^1$S$_0$ NN scattering is characterized by a large
negative scattering length indicating the presence of a nearly bound
resonant state at zero scattering energy. Around this low-energy
pole the NN interaction can be well approximated by a separable interaction
$v(k,k')\approx \lambda v(k) v(k')$ for which the so called inverse
scattering problem can be solved.
It means that both $v(k)$ as well as $\Delta (k_F)$ are in fact fully
determined by means of the phase shifts $\delta(k)$. Although, at first
glance, this approximation seems to be valid ({\it i\/}) only at low energies
and requires ({\it ii}) the knowledge of $\delta(k)$ at, in principle, all
energies it appears to work surprisingly well up to $k_F\sim 1.4$\,fm$^{-1}$.
Note however, that the effective range approximation to the phase
shifts works well only at low-densities up to $k_F\sim 0.6$\,fm$^{-1}$, see
Fig.~\ref{freegap}.

The inclusion of in-medium polarization and
screening corrections appear to be extremely difficult.
So far no consensus has been reached how to consistently compute
these corrections, see Refs.~\cite{[Lom01],[Dea03]} and refs. therein.
However, instead of deriving in-medium gap
equation $\Delta(k_F)$  from the bare NN interaction one can  attack
the problem starting directly from an effective interaction
like Gogny or local density dependent delta interaction (DDDI).
Indeed, as shown by Garrido {\it et al.\/}~\cite{[Gar99]} the pair gap
calculated using the
Gogny force with parameter set D1S fitted directly to finite
nuclei~\cite{[Ber84]} follows rather closely
the pair-gap calculated using the bare Paris-force
with Brueckner-Hartree-Fock spectrum. The difference between
the gaps increases with increasing density
reflecting, most likely, an enhancement due to the averaging over low-density
skin region which is implicitly taken into account through the
parameters of the effective Gogny interaction.
In Ref.~\cite{[Gar99]} it was demonstrated that
DDDI is also flexible enough to follow the Paris-force (and Gogny)
results, however, only after careful adjustment of the cutoff parameter.
A need for cutoff within the local pairing
theory is neither satisfactory nor
unambiguous and spoils to large extent its simplicity.  It is
interesting to observe, however, that local interactions
can be safely used in p-h channel
[Skyrme force] without any artificial cutoff and that
in fact in this channel it is equivalent, at least according
to effective theory principles, to the finite-range Gogny
interaction.

%//////////////////////////////////////////////////////////////////////////////
%//////////////////////////////////////////////////////////////////////////////

\section{Toward local superfluid effective theory.}\label{local}

A guiding principle underlying any effective theory aiming to
describe the low-energy limit of a deeper, more fundamental theory can be
formulated in the following way:
{\it The low-energy\/} (infrared) {\it phenomena are not sensitive
enough to resolve high-energy\/} (ultraviolet) {\it dynamics.\/}
It means that short-range (high-momentum) dynamics can be removed
from the theory and replaced (renormalized)
by a few local corrections. In momentum space one can therefore expand
a short-range (SR) interaction as:
\be\label{vq}
 v_{\text{SR}}(q^2) \approx g +  g_2 q^2  + g_4 q^4
 \ldots\, ,
 \ee
expressing it formally by means of few constants $g, g_2, g_4, \ldots$
which need to be carefully readjusted to a selected set of
low-energy data.  In coordinate representation interaction Eq.~(\ref{vq})
can be rewritten as:
 \bn\label{vcorr}
    v_{\text{SR}}({\boldsymbol r}) &\approx &
  c a^2 \delta_a ({\boldsymbol r}) \nonumber \\
 & + & d_1 a^4 {\boldsymbol \nabla}^2 \delta_a ({\boldsymbol r})
   +  d_2 a^4 {\boldsymbol \nabla} \delta_a ({\boldsymbol r})
   {\boldsymbol \nabla}\nonumber \\
 & + & \ldots \nonumber \\
 & + & h_1 a^{n+2} {\boldsymbol \nabla}^n \delta_a ({\boldsymbol r}) +\ldots
\, ,
\en
where $\delta_a (r)$ denotes an arbitrary model of the Dirac delta
function while $c,d_1,d_2,h_1,\ldots$ denote empirically adjustable
constants. In nuclear structure, due to the non-singular and very short range
nature of the effective interaction, one can in fact use the strict
limit
$\lim_{a\rightarrow 0} \delta_a ({\boldsymbol r})= \delta ({\boldsymbol r})$
which corresponds to the well-known Skyrme
interaction [first three terms in Eq.~(\ref{vcorr})]. On the other hand
by retaining
in Eq.~(\ref{vcorr}) only the first term modeled by a sum of attractive
and repulsive Gaussians of different ranges one obtains the well-known
finite-range Gogny force. Of course both forces must be augmented by the
density dependent and spin-orbit terms as well as by space-, spin-,
and isospin-exchange terms. The Skyrme  and Gogny forces are therefore two
realizations  [among infinitely many]
of effective nuclear interactions which are used in practical
nuclear structure calculations.

\begin{figure}[ht]
\begin{center}
%\includegraphics[scale=0.70, angle=0.0, clip, viewport= 150 300 600
%670]{so-fig01.eps}
\includegraphics[scale=0.50, angle=0.0,clip]{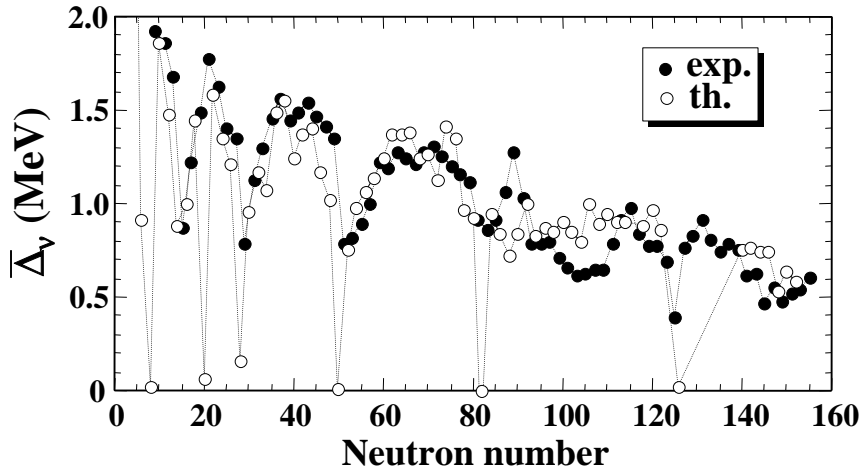}
\caption[]{Mean neutron pair gaps caclulated using HFB D1S Gogny
method (open dots) in Ref.~\protect{\cite{[Hil02]}} in comparison
to empirical three-point OES data $\Delta (odd-N)$ (filled dots).
}\label{gogny}
\end{center}
\end{figure}

The Gogny force can be unambiguously used also in the p-p
channel. Indeed, the finite-range ($r_o\sim 1$\,fm)
automatically discriminates states above $E_c\sim p_c^2/2m_r \sim
\hbar^2/mr_o \sim 40$\,MeV [$m_r=m/2$ is reduced mass]
since $r_o p_c\sim \hbar$.
The Gogny interaction was proved to be  very successful
in numerous practical applications. In particular,
the average Hartree-Fock-Bogolyubov (HFB) pair-gaps calculated using D1S
Gogny follows very accurately empirical three-point OES
$\Delta (N) \equiv (-1)^N [ B(N-1)+B(N+1) - 2B(N) ]/2$
calculated for odd-$N$, see Ref.~\cite{[Hil02]} and Fig.~\ref{gogny}.

%//////////////////////////////////////////////////////////////////////////////
%//////////////////////////////////////////////////////////////////////////////

Local delta-type or DDDI type pairing interactions are successfully
used particularly in conjunction with Skyrme interaction.
These applications explicitly invoke a cutoff parameter to
avoid divergences. It appears, however, that the divergence
can be rather easily identified and subsequently
regularized and a cutoff free local superfluid theory can be constructed.

To illustrate a possible regularization schemes
let us consider first the effective theory approach of Papenbrock and
Bertsch~\cite{[Pap99]}. In the contact-force approximation which
takes into account only first term in Eq.~(\ref{vq}) the BCS gap equation
takes the following form:
\be\label{gap}
1= -\frac{gV}{2(2\pi)^3} \int \frac{d^3
{\boldsymbol k}}{\sqrt{(\varepsilon_k - \lambda)^2 +\Delta^2} }\, ,
\ee where
$V$ and $\lambda$
denote volume and chemical potential, respectively. The integral in
Eq.~(\ref{gap}) is ultraviolet (high-momentum) divergent. It appears,
however, that the equation for scattering length, $a$, is also divergent
within contact approximation:
\be\label{scat}
-\frac{mgV}{4\pi a} +1 = -
\frac{gV}{2(2\pi)^3} \int \frac{d^3 {\boldsymbol k}}{\varepsilon_k}\, ,
\ee
and that the divergences are of the same type. Hence relation (\ref{scat})
can be used as a counter-term to regularize the gap equation:
\be\label{gap2}
\frac{m}{4\pi a}
= - \frac{1}{2(2\pi)^3} \int {d^3 {\boldsymbol k}} \left[
\frac{1}{\sqrt{(\varepsilon_k - \lambda)^2 +\Delta^2}} -
\frac{1}{\varepsilon_k} \right]\, .
\ee
This is a very elegant example of
regularization connecting the pair-gap (and contact-force strength
$g$) directly to the free two-particle scattering length.
However, the formalism applies only to dilute homogeneous medium.
Finite range corrections [i.e. higher order expansion terms in
Eq.~(\ref{vq})] are difficult to handle but can be, at least in principle,
systematically  implemented and regularized order by order.
Judging from Fig.~\ref{freegap} the lowest order  finite range
corrections are expected to extend the
validity of this scheme till $k_F\sim 0.6$\,fm$^{-1}$.

\smallskip

The problem of  ultraviolet type divergence in anomalous density matrix
persist also in finite nuclei:
\be
   \nu ({\boldsymbol r_1}, {\boldsymbol r_2}) =
   \sum_i v_i^*({\boldsymbol r_1}) u_i ({\boldsymbol r_2})
   \sim  \frac{1}{|{\boldsymbol r_1} - {\boldsymbol r_2} |}\, .
\ee
 Here the situation seems to be even more
complex since ({\it i\/}) realistic single-particle
(sp) spectra must be used right from the
beginning and ({\it ii\/})
it is not at all obvious what physical quantities
need to be used in order to regularize divergent terms.
The appropriate regularization scheme was proposed
recently by Bulgac and Yu~\cite{[Bul02],[Bul02a]}. Their scheme is
built upon the local density approximation (LDA) i.e.
takes automatically into account the dominant p-h channel.
The idea is to
introduce cutoff ($E_c \equiv \frac{(\hbar k_c)^2}{2m}$) dependent
counter-terms leading to standard local HFB formalism with cutoff
parameters but with a gap equation dependent on the effective
{\it running\/} coupling constant:
\bn
   \nu_c ({\boldsymbol r}) & = & \sum_{E_i \geq 0}^{E_c}
    v_i^*({\boldsymbol r}) u_i ({\boldsymbol r})\, ,    \\
   \Delta ({\boldsymbol r}) & = &  -g_{eff} ({\boldsymbol r})
   \nu_c ({\boldsymbol r}) \label{delta2}\, , \\
   \frac{1}{g_{eff} ({\boldsymbol r})} & = &
   \frac{1}{g[\rho ({\boldsymbol r})]} - \frac{m ({\boldsymbol r})
   k_c ({\boldsymbol r})}{2\pi^2 \hbar^2}
   \left\{ 1 -\frac{k_F ({\boldsymbol r})}{2k_c ({\boldsymbol r})}
   \mbox{ln}\frac{k_c ({\boldsymbol r})+k_F ({\boldsymbol r})}
                 {k_c ({\boldsymbol r})-k_F ({\boldsymbol r})}  \right\} .
\en
Introducing a
{\it running\/} coupling constant implies that the cutoff dependence is
only formal and disappears for sufficiently large
$E_c$~\cite{[Bul02],[Bul02a]}. This cutoff free superfluid LDA (SLDA)
approach is now in phase of extensive tests~\cite{[Bul02a],[Bul04]}.

\section{Entanglement}\label{ent}

%//////////////////////////////////////////////////////////////////////////////
\begin{figure}[t]
\begin{center}
%\includegraphics[scale=0.70, angle=0.0, clip, viewport= 150 300 600
%670]{so-fig01.eps}
\includegraphics[scale=0.50, angle=0.0,clip]{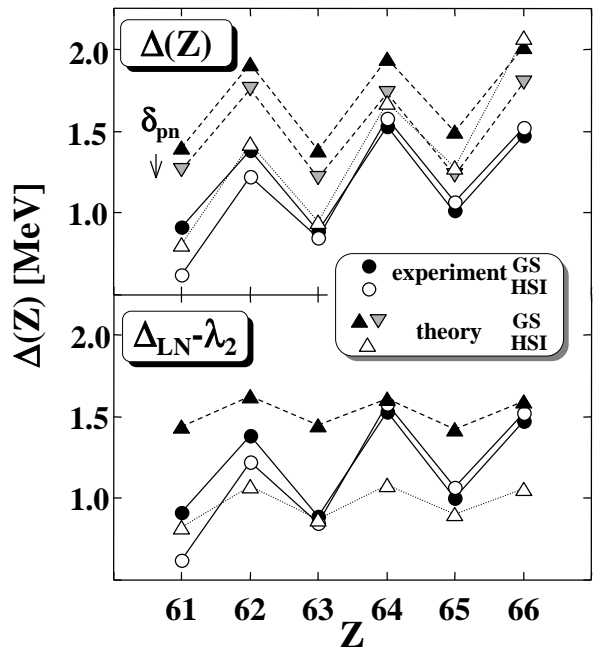}
\caption[]{Experimental OES (dots, solid lines) versus
calculated (triangles) OES (upper part) and
calculated value of mean proton LN gap $\Delta_{LN} - \lambda_2$
(lower part). Filled (open) symbols denote GS (HSI) values,
respectively. Gray triangles include theoretical GS values of
OES corrected by residual pn interaction.}
\label{oes}
\end{center}
\end{figure}
%//////////////////////////////////////////////////////////////////////////////

Pairing gaps are not directly accessible in experiment. Hence, various
indirect methods must be applied to extract information about them.
%Studies of odd-even staggering of nuclear binding energies (OES) or
%studies of moments of inertia at high-spins are just two prominent examples.
The major difficulty is that all these indirect methods
entangle pairing with various effects
including shape and shape-polarization effects, sp
splitting, time-odd fields, or beyond mean-field residual interaction effects
making life rather perplexing and, in fact, introducing in a natural way
uncertainties into our knowledge of nuclear pairing.
It is therefore desirable to hunt for simple physical
situations or phenomena where at least some of these {\it contaminants\/}
are either decoupled or can be relatively well controlled.
Superdeformation (SD) is one of the most prominent examples of such a
phenomenon. Indeed, stability of nuclear shape along the SD band
allows to study pairing correlations from the static to dynamic regime.
Let us recall that such concepts and techniques like
double-stretched quadrupole pairing~\cite{[Sat94a]}, the surface-active
DDDI~\cite{[Cha76]}, the Lipkin-Nogami (LN) number-projection~\cite{[Pra73]}
were applied for the first time in a {\it systematic way\/} in SD bands in
Hg-Pb nuclei~\cite{[Sat94a],[Gal94],[Val97],[Afa99]}. Afterwords
these methods became standard in large-scale calculations in
high-spin physics.

High-spin isomers (HSI) open yet another and so far unexplored
venue to study  pair correlations and blocking phenomena~\cite{[Dra98],[Xu98]}.
Thanks to their structural simplicity both configuration and
shape can be kept rather well under control. In turn,
shape and pairing polarization due to blocking
can be studied in detail and  traditional average gap
method~\cite{[Mol92]} used to determine pairing strength, $G_{MN}$,
can be re-examined. In particular, it was found that inclusion of these
polarization effects requires $\sim$10\%
larger $G$ as compared to $G_{MN}$~\cite{[Xu99]} to reproduce experimental OES.

Recently the HSI have been systematically observed in $N$=83 nuclei with
60$\leq$Z$\leq$67~\cite{[Oda05]}. This unique data set
enables to study for the first time OES both at the ground
states (GS) as well as at high-spins.
The most striking feature of this data set is the almost constant
excitation energy of the HSI what implies that
$\Delta_{GS}(Z)\approx \Delta_{HSI}(Z)$, see
Fig.~\ref{oes}. Conventional interpretation of this result
in terms of pairing-gap suggests the lack of
blocking phenomenon. The pair-gaps $\Delta_{LN} -
\lambda_2$ calculated using diabatic
Strutinsky type method involving self-consistent blocking and
LN particle-number projection~\cite{[Xu98],[Xu99]} shown
in lower part of Fig.~\ref{oes} clearly show that this conventional
interpretation is oversimplified. Contributions to OES from
pairing (blocking), sp-proton energy splitting and
residual proton-neutron (pn) interaction must all be taken into account
to reproduce experimental data in a
satisfactory way, see upper part of Fig.~\ref{oes}.
Even then, contribution due to blocking is clearly too
strong and requires revisiting.

\section{Hunting for fingerprints of isoscalar pn pairing
collectivity.}\label{pnpair}

%//////////////////////////////////////////////////////////////////////////////
\begin{figure}[t]
\begin{center}
%\includegraphics[scale=0.70, angle=0.0, clip, viewport= 150 300 600
%670]{so-fig01.eps}
\includegraphics[scale=0.50, angle=0.0,clip]{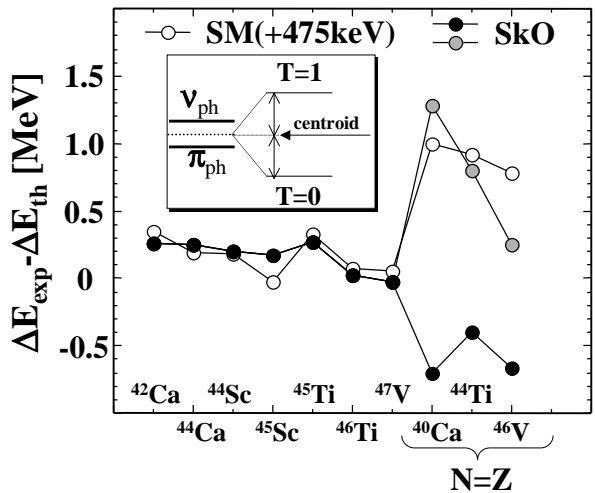}
\caption[]{Calculated energy differences $\Delta E_{exp} - \Delta E_{th}$
between  terminating states,
$\Delta E = E(d_{3/2}^{-1} f_{7/2}^{n+1}) - E(f_{7/2}^n)$,
in A$\sim$50 mass region relative to experimental data.
Open symbols denote state of the art SM results
(shifted by 475keV) while
filled symbols denote the SHF calculations using the SkO
parameterization with spin-orbit term reduced by 5\%.
Gray dots denote the SHF results in $N$=$Z$ nuclei corrected
phenomenologically for isospin breaking
effect, see insert and Ref.~\protect{\cite{[Sto05]}} for more details.
}\label{termin}
\end{center}
\end{figure}
%//////////////////////////////////////////////////////////////////////////////

The existence of isoscalar ($t=0$) pairing phase in
atomic nuclei is still a fascinating open issue. Various phenomena
are discussed in this context including an onset of $t=0$ pairing
driven by nuclear rotation. However, neither the suggested shifts in
ground-band S-band crossing frequency nor
substantial changes in moment of inertia were convincingly
confirmed by experiment. On the contrary standard calculations
seem to work reasonably good in particular
in the most promising area of heavy deformed $A$$\sim$80, $N\sim Z$
nuclei~\cite{[Rud97],[Afa05]}. There are, however, new data
indicating rather unusual band crossing phenomena
in e.g. $^{73}$Kr~\cite{[Kel02]}. Indeed, standard
mean-field calculations~\cite{[Kel02]} can reproduce the $^{73}$Kr data assuming
two different structures below [one-quasi-particle (QP)
band built upon negative parity Nilsson-level originating from
$\nu (f_{5/2}$-$p_{3/2})$ subshell] and
above back-bending where odd-neutron occupy positive-parity Nilsson state
originating from $\nu g_{9/2}$ subshell while protons
form 2QP structure involving $\pi (f_{5/2}$-$p_{3/2})\otimes \pi g_{9/2}$.
Strong E2 transitions connecting these structures
({\it i\/}) cannot be explained within standard mean-field calculations and
({\it ii\/}) indicate unusually strong configuration mixing.
Whether or not this configuration mixing can be accounted for within
mean-field approximation invoking $t=0$ pn-pairing i.e. novel spontaneous
symmetry breaking mechanism resulting in pn-mixing
is under study~\cite{[Wys05]}.

Also calculated energy differences $\Delta E = E(d_{3/2}^{-1}
f_{7/2}^{n+1}) - E(f_{7/2}^n)$ between band-terminating states in A$\sim$50
seem to provide new evidence for $t=0$ pairing~\cite{[Sto05]}. As shown in
Fig.~\ref{termin} the values of $\Delta E_{exp} - \Delta E_{th}$ calculated
using state of the art
shell-model (SM) and Skyrme-Hartree-Fock (SHF) method
follow, up to a constant offset of $\sim$475\, keV, very closely
each other. The quantitative difference between
the accuracy of theoretical predictions in $N \ne Z$ as compared
to $N=Z$ nuclei seen both in the SM as well as in the SHF
calculations suggests, most
likely, an enhanced $t=0$ pair scattering from $sd$ to $fp$
shell which is beyond
present shell-model space. Similar effect was suggested in
Ref.~\cite{[Ter98]}.

One of the most promising signal of $t=0$ pairing comes from binding
energies in $N=Z$ nuclei, where the problem of the Wigner
energy (WE) [extra binding energy]
is known to plague mean-field masses, see Ref.~\cite{[Lun03]}
and refs. therein. It is relatively well established from nuclear
shell-model studies
that the WE is predominantly due to $t=0$ interaction~\cite{[Bre90],[Sat97a]}.
The SM indicates, however, that the structure of the WE is very complex and is
not dominated neither by $L=0,S=1,T=0$~\cite{[Pov98]} nor by
$J=1,T=0$~\cite{[Sat97a]} isoscalar pn-pairs.
Within mean-field model, which is entirely based upon the concept
of spontaneous symmetry breaking, the definition of $t=0$ pairing
in terms of symmetry conserving pairing interaction consisting
only $L=0,S=1,T=0$ (or $J=1,T=0$) pairs is not at all justified.
Although the form of effective $t=0$ pn-pairing interaction appropriate
for mean-field calculations is an open issue it is
rather well established qualitatively that
mean-field model augmented by $t=0$ pairing is capable to heal the
mass-defect problem around $N\sim Z$ line~\cite{[Sat97],[Eng96],[Rop00]}.
Quantitative estimate of $t=0$ pairing within mean-field requires, however,
a reliable evaluation of corrections due to isospin-symmetry restoration
which go beyond mean-field. Within random-phase-approximation these
corrections roughly restore linear term in the nuclear symmetry energy
(NSE) giving rise to $\sim T(T+1)$ dependence of the NSE~\cite{[Nee03x]}.
It appears however that at least part of the linear term ($\sim T$) is
incorporated already at the level of (self-consistent) mean-field.
Hence, further progress in the field
is impossible without thorough understanding of the NSE
being currently under intense studies~\cite{[Cha03],[Sat03],[Ban05],[Sat05a]}.

\bigskip

This work has been supported by the Polish Committee for Scientific
Research (KBN) under contract 1~P03B~059~27.

%\bibliography{rev}

\begin{thebibliography}{10}

\bibitem{[Boh58]}
{A. Bohr, B.R. Mottelson, and D. Pines, Phys. Rev. {\bf 110} (1958) 936.}

\bibitem{[Bel59]}
{S.T. Belyaev, Mat. Fys. Medd. Dan. Vid. Selsk. {\bf 31} (No. 11) (1959).}

\bibitem{[Nil61]}
{S.G. Nilsson and O. Prior, Mat. Fys. Medd. Dan. Vid. Selsk. {\bf 32} (No. 16)
  (1961).}

\bibitem{[Sat05]}
{W. Satu{\l}a and R. Wyss, Rep. Prog. Phys. {\bf 68} (2005) 131.}

\bibitem{[Elg98]}
{{\O}. Elgar{\o}y and M. Hjorth-Jensen, Phys. Rev. {\bf C57} (1998) 1174.}

\bibitem{[Lom01]}
{U. Lombardo and H.J. Schultze, {\it Lecture Notes in Physics\/} (Springer, New
  York) Vol. 578, 2001, p. 30.}

\bibitem{[Dea03]}
{D.J. Dean and M. Hjorth-Jensen, Rev. Mod. Phys. {\bf 75} (2003) 607.}

\bibitem{[Gar99]}
{E. Garrido {\it et al.\/}, Phys. Rev. {\bf C60} (1999) 064312.}

\bibitem{[Ber84]}
{J.-F. Berger, M. Girod, and D. Gogny, Nucl. Phys. {\bf A428} (1984) 23c.}

\bibitem{[Hil02]}
{S. Hilaire {\it et al.\/}, Phys. Lett. {\bf B531} (2002) 61.}

\bibitem{[Pap99]}
{T. Papenbrock and G.F. Bertsch, Phys. Rev. {\bf C59} (1999) 2052.}

\bibitem{[Bul02]}
{A. Bulgac and Yongle Yu, Phys. Rev. Lett. {\bf 88} (2002) 042504.}

\bibitem{[Bul02a]}
{A. Bulgac, Phys. Rev. {\bf C65} (2002) 051305.}

\bibitem{[Bul04]}
{A. Bulgac and Yongle Yu, Int. J. Mod. Phys. {\bf E13} (2004) 147.}

\bibitem{[Sat94a]}
{W. Satu{\l}a and R. Wyss, Phys. Rev. {\bf C50} (1994) 2888.}

\bibitem{[Cha76]}
{R.R. Chasman, Phys. Rev. {\bf C14} (1976) 1935.}

\bibitem{[Pra73]}
{H.C. Pradhan, Y. Nogami and J. Law, Nucl. Phys. {\bf A201} (1973) 357}.

\bibitem{[Gal94]}
{B. Gall {\it et al.\/}, Z. Phys. {\bf A348} (1994) 183.}

\bibitem{[Val97]}
{A. Valor, J.L. Egido, and L.M. Robledo, Phys. Lett. {\bf B392} (1997) 249.}

\bibitem{[Afa99]}
{A.V. Afanasjev, J. K\"onig, and P. Ring, Phys. Rev. {\bf C60}, R051303
  (1999).}

\bibitem{[Dra98]}
{G.D. Dracoulis, F.G. Kondev, and P.M. Walker, Phys. Lett. {\bf B419} (1998)
  7.}

\bibitem{[Xu98]}
{F.R. Xu, P.M. Walker, J.A. Sheikh, and R. Wyss, Phys. Lett. {\bf B435} (1998)
  257.}

\bibitem{[Mol92]}
{P. M\"oller and J.R. Nix, Nucl. Phys. {\bf A536} (1992) 20.}

\bibitem{[Xu99]}
{F.R. Xu, R. Wyss, and P.M. Walker, Phys. Rev. {\bf C60} (1999) 051301.}

\bibitem{[Oda05]}
{A. Odahara {\it et al.\/}, submitted to Phys. Rev. Lett. (2005).}

\bibitem{[Sto05]}
{G. Stoicheva {\it et al.\/}, in preparation (2005).}

\bibitem{[Rud97]}
{D.~Rudolph {\it et. al.}, Phys. Rev. {\bf C}56 (1997) 98.}

\bibitem{[Afa05]}
{A.V. Afanasjev and S. Frauendorf, Phys. Rev. {\bf C71} (2005) 064318.}

\bibitem{[Kel02]}
{N.S. Kelsall {\it et al.\/}, Phys. Rev. {\bf C65} (2002) 044331.}

\bibitem{[Wys05]}
{R. Wyss and W. Satu{\l}a, in preparation (2005)}.

\bibitem{[Ter98]}
{ J.~Terasaki, R.~Wyss, and P.-H.~Heenen, Phys. Lett. {\bf B437} (1998) 1.}

\bibitem{[Lun03]}
{D. Lunney, J.M. Pearson, and C. Thibault, Rev. Mod. Phys. {\bf 75} (2003)
  1021.}

\bibitem{[Bre90]}
{D.S. Brenner {\it et al.\/}, Phys. Lett. {\bf 243B} (1990) 1.}

\bibitem{[Sat97a]}
{W. Satu{\l}a {\it et al.\/}, Phys. Lett. {\bf B407} (1997) 103.}

\bibitem{[Pov98]}
{A. Poves and G. Mart\'inez-Pinedo, Phys. Lett. {\bf B430} (1998) 203.}

\bibitem{[Sat97]}
{W. Satu{\l}a and R. Wyss, Phys. Lett. {\bf B393} (1997) 1.}

\bibitem{[Eng96]}
{J. Engel, K. Langanke, and P. Vogel, Phys. Lett. {\bf B389} (1996) 211.}

\bibitem{[Rop00]}
{G. R\"opke {\it et. al.\/}, Phys. Rev. {\bf C61} (2000) 024306.}

\bibitem{[Nee03x]}
{K. Neerg{\aa}rd, Phys. Lett. {\bf B537} (2002) 287; {\bf B572} (2003) 159.}

\bibitem{[Cha03]}
{R.R. Chasman, Phys. Lett. {\bf B577} (2003) 47.}

\bibitem{[Sat03]}
{W. Satu{\l}a and R. Wyss, Phys. Lett. {\bf B572} (2003) 152.}

\bibitem{[Ban05]}
{S. Ban, J. Meng, W. Satu{\l}a, and R. Wyss, submitted to Phys. Lett. {\bf B}}.

\bibitem{[Sat05a]}
{W. Satu{\l}a, R. Wyss, and M. Rafalski, nucl-th/0508004, submitted to Phys.
  Rev. Lett. (2005).}

\end{thebibliography}
%\bibliographystyle{unsrt}

\end{document}